\documentclass[10pt, eps, twocolumn,showpacs,pre]{revtex4-1}
\usepackage{url,ulem}
\usepackage{fourier}
\usepackage{amsmath,amssymb}
\usepackage{graphicx,color}
\usepackage{setspace}

\newcommand{\dfracp}[2]{\dfrac{\partial #1}{\partial #2}}
\newcommand{\ave}[1]{\left\langle #1 \right\rangle}

\begin{document}
\title{Conditions for predicting
  quasistationary states by rearrangement formula}
\author{Yoshiyuki Y. Yamaguchi}
\email[E-mail: ]{yyama@amp.i.kyoto-u.ac.jp}
\affiliation{
	Department of Applied Mathematics and Physics, 
	Graduate School of Informatics, Kyoto University, 
	606-8501 Kyoto, Japan}
\author{Shun Ogawa}
\email[E-mail: ]{shun.ogawa@cpt.univ-mrs.fr}
\thanks{On leave from 
Department of Applied Mathematics and Physics, Kyoto University.}
\affiliation{
	Aix-Marseille Universit\'e, Universit\'e de Toulon,
	CNRS, Centre de Physique Th\'eorique UMR 7332, 13288 Marseille Cedex 9, France
	}

\begin{abstract}
    Predicting the long-lasting quasistationary state for a given initial state
    is one of central issues in Hamiltonian systems having long-range interaction.
    A recently proposed method is based on the Vlasov description
    and uniformly redistributes the initial distribution
    along contours of the asymptotic effective Hamiltonian,
    which is defined by the obtained quasistationary state and
    is determined self-consistently.
    The method, to which we refer
    as the rearrangement formula,
    was suggested to give precise prediction under limited situations.
    Restricting initial states 
    consisting of spatially homogeneous part and small perturbation, 
    we numerically reveal two conditions
      that the rearrangement formula prefers:
    One is no Landau damping condition for unperturbed 
    homogeneous part,
    and the other comes from the Casimir invariants.
    Mechanisms of these conditions are discussed.
    Clarifying these conditions,
    we inform validity to use the rearrangement formula
    as the response theory for an external field,
    and we shed light on improving the theory
    as a nonequilibrium statistical mechanics.
\end{abstract}
\maketitle 

\section{Introduction}

Long-range interaction violates some assumptions
introduced in the equilibrium statistical mechanics 
and thermodynamics,
for instance additivity \cite{AC14, AC09}.
One remarkable phenomenon in such a system is existence
of long-lasting nonequilibrium quasistationary states (QSSs)
in the relaxation process, and the life time of QSSs diverges 
in the limit of large population \cite{JB87, ZM03, YYY04, MK11}.
In the mean-field limit,
dynamics of the system is described
by the Vlasov equation (or collisionless Boltzmann equation)
\cite{BH77,Dobrushin79,Spohn},
and QSSs are regarded as stable stationary solutions
to the Vlasov equation.
QSSs are said to be found in various scales in the nature,
from the laboratory scale as the plasma crystals \cite{FCFD95,SDFD99, KN06}
to extremely large scale as the elliptic or spherical galaxies \cite{JB87}.
A central issue of long-range interacting systems
is to predict the QSS from a given nonstationary initial state. 

One theoretical approach is proposed by Lynden-Bell \cite{DLB67},
which is originally proposed for the self-gravitating systems
and is easy to use for the so-called waterbag initial states.
Several tests are performed for the theory in the self-gravitating systems
with 1D \cite{YYY08, MJ11}, 2D \cite{TNT10} and 3D \cite{LPR08},
and the Hamiltonian mean-field (HMF) model (or the globally coupled XY model)
\cite{FPCB12}.
In the self-gravitating systems, thanks to homogeneity of potential,
initial states are classified by the virial ratio,
and the Lynden-Bell's theory gives good prediction of QSSs
if initial states satisfy the virial condition.
The concept of virialization is extended for nonhomogeneous potential
of the HMF model \cite{FPCB12}
to avoid parametric resonance making halo \cite{RLG94,RP11}.
The generalized virial condition helps to prepare
initial states for which QSSs are described by the Lynden-Bell's theory.
See also Ref. \cite{YL14}.

Another approach is the rearrangement formula,
or the integrable (uncoupled) model.
In this article we consider asymptotic states of the Vlasov dynamics
which are QSSs in which we are interested.
The idea to get the asymptotic state
is to redistribute the initial distribution
along contours of the asymptotic effective Hamiltonian,
which is determined by the asymptotic state,
and to solve the self-consistent equation for the asymptotic state.
The rearrangement formula is introduced without theoretical justification,
but is successfully examined in the HMF model 
for single-level \cite{XL09,BMR11} and multi-level \cite{ACRT14}
waterbag initial states numerically. 
Further, the theory also gives good prediction for 3D self-gravitating systems 
for the waterbag initial states, and for the parabolic initial states \cite{BRPL14}.

The two different theories of the rearrangement formula
and of Lynden-Bell prefer the generalized virial states,
but the former is said to provide more accurate predictions than the latter
\cite{ACRT14,BRPL14}.
We then focus on the rearrangement formula rather
than the Lynden-Bell's theory.
Another reasoning to focus on the rearrangement formula
is that the formula is useful even for non-waterbag initial states. 
Indeed, for perturbed states from stable stationary states,
disordered thermal equilibria for instance,
the rearrangement formula is theoretically justified \cite{SO14}
by use of the asymptotic-transient field decomposition
and the transient (T-)linearized Vlasov equation \cite{CL98,CL03,CL09}.
We note that the naming is ``linearization'',
but the equation includes a nonlinear term as shown later.
A similar formula is also derived via the variational principle
in the context of plasma waves \cite{IYD11,IYD14}.
However, as the Lynden-Bell's theory,
the rearrangement formula is not always precise
and a previous work \cite{SO14} suggests that the stable stationary state
with zero Landau damping \cite{LDL46} rate is preferred
as the unperturbed states.
We refer to this condition as the no Landau damping condition.

It is still unclear if the no Landau damping condition
is more relevant than the virial condition, and if the former
is solely essential,
since numerical tests have been performed
for a limited situation.
Moreover, the no Landau damping condition is for the unperturbed states,
and hence one may expect a condition for the whole initial state
including perturbation.
The main purposes of the present article are
to confirm the relevance of the no Landau damping condition
and to reveal one more condition relating to the Casimir invariance
by performing systematic numerical simulations
of the Vlasov equation.

It is important to clarify validating conditions
of the rearrangement formula from the following two contexts.
One is as the response theory for external field.
The rearrangement formula gives non-classical critical exponents,
and the theoretical predictions are in good agreement
with numerical simulations \cite{OPY14,SO14,OY15}. 
We can further justify the non-classical critical exponents
by showing that the validating conditions are satisfied
for computing the response.
The other is related to improvement of the theory.
After confirming the validating conditions, 
it might be possible to improve the theory
by including the Landau damping into the rearrangement formula,
for instance.

This article is organized as follows.
We first introduce the HMF model and the associated Vlasov equation
in Sec.~\ref{sec:model}.
In Sec.~\ref{sec:rearrangement-formula},
we briefly review the rearrangement formula,
and give theoretical predictions for the HMF model.
The section \ref{sec:numerics} is for examinations of the two conditions: 
In Sec.~\ref{sec:two-conditions} we explain why one may expect
the conditions.
The no Landau damping condition for the reference state
is carefully confirmed in Sec.~\ref{sec:no-Landau-damping},
and a new condition of the Casimir invariance is reported
in Sec.~\ref{sec:Casimir}.
Based on the numerical findings,
we discuss validity for using the rearrangement formula
as the response theory to the external force,
in particular to compute the critical exponents, 
in Sec. \ref{eq:response-theory}.
The final section \ref{sec:summary} is devoted to a summary and discussions.

\section{The Hamiltonian mean-field model
and the Vlasov equation}
\label{sec:model}

The HMF model \cite{IK93,MA95} is a model of a ferromagnetic body,
and is expressed by the Hamiltonian 
\begin{equation}
  \begin{split}
    \label{eq:HN}
    H_{N} = \sum_{i=1}^N \frac{p_{i}^{2}}{2}
    +& \dfrac{1}{2N} \sum_{i,j=1}^N [1-\cos(q_{i} -q_{j})] \\
    &- \sum_{i=1}^{N} \left(h_{x}(t)\cos q_{i} + h_{y}(t)\sin q_{i}\right),
  \end{split}
\end{equation}
where the last two terms express 
the interaction energy between XY-spins (rotators) and
the external magnetic field $(h_{x}(t),h_{y}(t))$.
Response to the external field will be discussed
in Sec.\ref{eq:response-theory},
and until then, the external field is set as zero. 
The system is also looked on as a dynamical system with 
many particles moving on the unit circle with attractive all-to-all interactions,
and the position and the conjugate momentum of $i$-th particle are
denoted by $q_{i}$ and $p_{i}$ respectively
defined in $q_{i}\in (-\pi,\pi]$ and $p_{i}\in \mathbb{R}$.
The HMF model is a paradigmatic toy model,
and the simple interaction provides advantages in theory and in numerics.

When one takes the limit of $N\to \infty$,
temporal evolution of the HMF model can be well described
in terms of the single particle distribution $f(q,p,t)$
governed by the Vlasov equation \cite{BH77,Dobrushin79,Spohn} 
\begin{equation}
    \label{eq:Vlasov}
    \partial_{t} f + \left\{\mathcal{H}[f],f\right\} = 0,\quad f(q,p,0) = f_{\rm I}(q,p),
\end{equation}
where the Poisson bracket $\{a,b\}$ is defined as 
\begin{equation}
  \{ a, b \} = \dfracp{a}{p} \dfracp{b}{q} - \dfracp{a}{q} \dfracp{b}{p} 
\end{equation}
for two functions on the $\mu$ space $(-\pi,\pi]\times\mathbb{R}$.
The effective Hamiltonian $\mathcal{H}[f]$ is given by 
\begin{equation}
  \label{eq:EH}
  \mathcal{H}[f] = p^{2}/2  + \mathcal{V}[f](q,t)
\end{equation}
with
\begin{equation}
  \mathcal{V}[f] = - ( \mathcal{M}_{x}[f] + h_{x} ) \cos q
  - ( \mathcal{M}_{y}[f] + h_{y} ) \sin q.
\end{equation}

In this article, we look into the dynamics
through the magnetization (or the order parameter) vector
$(\mathcal{M}_{x}[f],\mathcal{M}_{y}[f])$ defined by
\begin{equation}
    \mathcal{M}_x[f]+ i\mathcal{M}_y[f] = \iint e^{iq} f(q,p,t){\rm d}q{\rm d}p.
\end{equation}
The magnetization vector has the modulus less than or equal to $1$,
and measures how particles concentrate at a certain direction
on the circle. If particles are uniformly distributed,
then $\mathcal{M}_{x}[f]=\mathcal{M}_{y}[f]=0$.
If particles are squeezed at a point on the unit circle, for instance $q=0$,
then $\mathcal{M}_{x}[f]=1$ and $\mathcal{M}_{y}[f]=0$.

\section{Rearrangement formula}
\label{sec:rearrangement-formula}

\subsection{General derivation}

We shortly review the rearrangement formula
in the absence of the external field.
See \cite{SO14} for theoretical justification of the formula.

We start from the initial state $f_{\rm I}$ close
to a spatially homogeneous stable stationary state $f_{\rm S}$, 
and decompose it into the two parts as
\begin{equation}
    \label{eq:fI}
    f_{\rm I}(q,p) = f_{\rm S}(p) + \epsilon g_{\rm I}(q,p), 
    \quad
    |\epsilon|\ll 1.
\end{equation}
We note that we can construct the rearrangement formula
even if the unperturbed part is spatially inhomogeneous,
but we restrict ourselves to the homogeneous case for simplicity.
Our interest is to predict the asymptotic state 
of the Vlasov dynamics denoted by $f_{\rm A}$,
which is assumed to be stationary.
It should be noted that the perturbed state \eqref{eq:fI}
possibly does not go to a stationary state, 
but to an oscillatory state
by forming small traveling clusters 
under some conditions \cite{JB09}.
We do not look into such states in the present article.

One standard method to analyze dynamics around $f_{\rm S}$
is to linearize the Vlasov equation \eqref{eq:Vlasov}
by expanding $f$ into
\begin{equation}
  f(q,p,t) = f_{\rm S}(p) + \epsilon g(q,p,t).
\end{equation}
The linearized Vlasov equation,
\begin{equation}
  \label{eq:LVE}
  \partial_{t} g + \{ \mathcal{H}[f_{\rm S}], g \} + \{ \mathcal{V}[g], f_{\rm S} \}
  = 0,
\end{equation}
gives the well-known Landau damping \cite{LDL46} of perturbation.
If the Landau damping is strong enough,
then the asymptotic state
$f_{\rm A}$ may coincide with the initial stable stationary reference $f_{\rm S}$.
On the other hand,
if the Landau damping rate is close to zero,
then nonlinear trapping \cite{TO65}
stops the damping, and
the system relaxes to a different asymptotic state from $f_{\rm S}$.
In the latter case,
due to the nonlinearity,
predicting the asymptotic state is nontrivial.
The rearrangement formula is a powerful tool 
in the latter case as shown in this article.

The key idea of the rearrangement formula is as follows.
Imagine that the initial state $f_{\rm I}$ asymptotically goes to
a stationary state $f_{\rm A}$, which is still unknown.
The asymptotic state constructs
the asymptotic effective Hamiltonian $\mathcal{H}_{\rm A}=\mathcal{H}[f_{\rm A}]$
of the Vlasov equation,
and the asymptotic Hamiltonian drives the system
for a long time and takes it to the asymptotic state $f_{\rm A}$.
Then, we check self-consistency
between the imagined asymptotic state
and the driven asymptotic state.

The above idea is theoretically
formulated as follows.
We decompose $f$ into another way as
\begin{equation}
  \label{eq:ATP}
  f(q,p,t)= f_{\rm A}(q,p) + \epsilon g_{\rm T}(q,p,t) ,
\end{equation}
where $f_{\rm A}$ and $\epsilon g_{\rm T}$ are respectively called 
the asymptotic (A-) and (T-)parts.
The A-part $f_{\rm A}$ is picked up by use of a special case
of the Abel's formula, 
\begin{equation}
  \label{eq:AP}
  f_{\rm A}(q,p)
  = \lim_{T\to \infty} \frac{1}{T} \int_0^T f(q,p,t) {\rm d}t, 
\end{equation}
and coincides with $\lim_{t \to \infty} f(q,p,t)$ if it exists.
According to the A-T decomposition \eqref{eq:ATP}, the effective
Hamiltonian is similarly decomposed as
\begin{equation}
    \label{eq:ATH}
        \mathcal{H}[f] = \mathcal{H}_{\rm A} + \epsilon\mathcal{V}_{\rm T},    
\end{equation}
where the A-part and the T-part are defined by
\begin{equation}
    \mathcal{H}_{\rm A}= p^{2}/2 + \mathcal{V}[f_{\rm A}],
    \quad
    \mathcal{V}_{\rm T} = \mathcal{V}[g_{\rm T}]
\end{equation}
respectively. 
Substituting the decomposition \eqref{eq:ATH}
into the Vlasov equation \eqref{eq:Vlasov}, we have
\begin{equation}
  \label{eq:ATVlasov}
  \partial_{t}f + \{\mathcal{H}_{\rm A}, f\}
  + \epsilon\{ \mathcal{V}_{\rm T}, f\} = 0. 
\end{equation}
If $f$ is always in an $O(\epsilon)$ neighborhood of $f_{\rm I}$,
we can approximate the above exact equation as
\begin{equation}
  \label{eq:TLVlasov}
  \partial_{t} f + \{\mathcal{H}_{\rm A}, f\}
  + \epsilon\{\mathcal{V}_{\rm T}, f_{\rm I}\} =0
\end{equation}
by omitting $O(\epsilon^{2})$ term
which couples with the T-part
$\mathcal{V}_{\rm T}$.
Nevertheless, we emphasize that the approximated equation
\eqref{eq:TLVlasov} is not just a linearized equation like Eq. \eqref{eq:LVE} 
for $O(\epsilon)$ terms,
since $f$ includes the $O(1)$ term of $f_{\rm A}$.
In other words,
the term $\{\mathcal{H}_{\rm A}, f\}$ has nonlinearity.
We remark that the criteria of truncation concerns to the surviving
time scale of each term \cite{CL98, CL03, CL09}.

We can show that, under some assumptions, 
the unknown transient field $\mathcal{V}_{\rm T}$
appearing in the third term of the left-hand-side of Eq. \eqref{eq:TLVlasov}
does not contribute to determine the effective Hamiltonian $\mathcal{H}_{\rm A}$ \cite{SO14}.
Therefore, {\it roughly} speaking,
the A-part $f_{\rm A}$ is obtained as the asymptotic solution
to the reduced equation
\begin{equation}
  \label{eq:AVlasov}
  \partial_{t} f + \{\mathcal{H}_{\rm A}, f\} =0. 
\end{equation}
Temporal evolution of $f$ is, hence, obtained as
\begin{equation}
  f(q,p,t) = f_{\rm I}(q_{-t},p_{-t}),
\end{equation}
where $(q_{t},p_{t})$ is the orbit of the Hamiltonian dynamics
governed by $\mathcal{H}_{\rm A}$ with the initial condition $(q,p)$.
The Abel's formula \eqref{eq:AP} gives
\begin{equation}
  \label{eq:AP-fA}
  f_{\rm A}(q,p)
  = \lim_{T\to \infty} \frac{1}{T} \int_0^T f_{\rm I}(q_{-t},p_{-t}) {\rm d}t,
\end{equation}
and this is the time average of $f_{\rm I}$ along the orbit
of the integrable Hamiltonian system $\mathcal{H}_{\rm A}$.
Thus, introducing the angle-action variables $(\theta,J)$
associated with $\mathcal{H}_{\rm A}$,
which is written as a function of $J$ only as $\mathcal{H}_{\rm A}(J)$,
the ergodic-like formula replaces the time average of Eq. \eqref{eq:AP-fA}
with the iso-$J$ average 
\begin{equation}
    \label{eq:fA}
    f_{\rm A} = \dfrac{1}{2\pi} \int_{-\pi}^{\pi}
    f_{\rm I}(q(\theta,J),p(\theta,J)) {\rm d}\theta
    = \ave{f_{\rm I}}_{\mathcal{H}_{\rm A}}.
\end{equation}
This expression \eqref{eq:fA} is the rearrangement formula,
on which we will discuss.
The concrete forms of $(\theta,J)$ and
$\ave{\bullet}_{\mathcal{H}_{\rm A}}$ is exhibited in the Appendix \ref{sec:explicit-average}
with another equivalent practical expression of the average.

\begin{figure}[tb]
  \begin{center}
    \includegraphics[width=8.5cm]{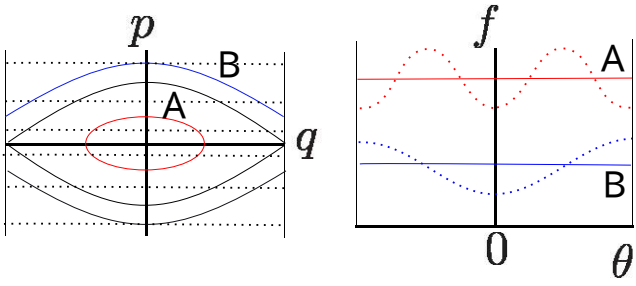}
    \caption{
      (color online)
     	 Schematic picture for the rearrangement formula \eqref{eq:fA}
     	 in the HMF model. 
         The left panel shows $\mu$ space,
         and dotted and solid lines represent
         contours of $f_{\rm I}$ and $f_{\rm A}$ respectively.
         The right panel shows $\theta$ dependence of distribution functions,
         where the angle variable $\theta$ is defined by the asymptotic
           effective Hamiltonian $\mathcal{H}_{\rm A}$,
         and the red solid line marked by A
         corresponds to the red solid contour A of the left panel.
         Along the contour, the initial state $f_{\rm I}$ depends
         on the angle $\theta$ as described by the red dotted line
         in the right panel.
         The blue solid contour marked by B is another example.
         For the two examples A and B,
         the positive $p$-axis corresponds to $\theta=0$.
           In this schematic picture $f_{\rm I}$ is assumed
           to be spatially homogeneous and a decreasing function of energy,
           though the rearrangement formula is also applicable to a spatially inhomogeneous $f_{\rm I}$.
    }
    \label{fig:rearrangement-formula}
  \end{center}
\end{figure}

An illustrative presentation 
of the rearrangement formula \eqref{eq:fA} is 
to redistribute height of the initial state $f_{\rm I}$ uniformly
on each contour of the asymptotic Hamiltonian $\mathcal{H}_{\rm A}$
as described in Fig. \ref{fig:rearrangement-formula}.
This procedure is consistent with the Jeans theorem \cite{JHJ15}
for constructing a stationary state,
since the resulting state is constant on each contour of $\mathcal{H}_{\rm A}$.
We note that neither the asymptotic state $f_{\rm A}$
nor Hamiltonian $\mathcal{H}_{\rm A}$ are still known,
since both sides of the formula \eqref{eq:fA}
depend on the asymptotic state.
We have to determine the asymptotic state
as it satisfies the self-consistent equation,
and the determination will be done in the next subsection \ref{sec:HMF-case}
for the HMF model.

It might be worth noting similarity between
the rearrangement formula and the Lynden-Bell's theory.
In the latter, we consider a waterbag initial state
and divide the phase space into small phase space elements.
Due to imcompressibility of the Vlasov flow,
the phase space elements are exclusive,
and hence we redistribute them to phase space
as maximizing the Fermi-Dirac like entropy
with keeping the invariants of mass, momentum, and energy.
In the former, we redistribute the phase space elements
as the Lynden-Bell's theory, but the redistribution is performed
on each iso-$J$ contour.

\subsection{Application to the Hamiltonian mean-field model}
\label{sec:HMF-case}

From symmetry of the system, we may assume that $\mathcal{M}_{y}=0$.
We determine the asymptotic Hamiltonian without the external field,
\begin{equation}
  \mathcal{H}_{\rm A} = p^{2}/2 - M_{\rm A} \cos q,
\end{equation}
by solving the self-consistent equation
\begin{equation}
    \label{eq:self-consistent}
    M_{\rm A} = \iint \cos q \ave{f_{\rm I}}_{\mathcal{H}_{\rm A}} {\rm d}q{\rm d}p
    = \iint \ave{\cos q}_{\mathcal{H}_{\rm A}} f_{\rm I} {\rm d}q{\rm d}p.
\end{equation}
In the last equality we used the fact that ${\rm d}q{\rm d}p={\rm d}\theta {\rm d}J$.
The self-consistent equation \eqref{eq:self-consistent} always has
the solution of $M_{\rm A}=0$ corresponding to the strong Landau damping case,
but we skip this trivial solution.
The non-zero solution of $M_{\rm A}$, solved numerically,
is the theoretical prediction to be examined in Sec.\ref{sec:numerics}.
Before going to numerical tests,
we observe theoretically obtained approximate solutions to self-consistent equation
by expanding the self-consistent equation \eqref{eq:self-consistent}
with respect to small $M_{\rm A}$ \cite{SO14}.
The expansion leads
\begin{equation}
  \label{eq:MA-expanded}
    A[f_{\rm I}] M_{\rm A}^{1/2}
    + D[f_{\rm I}] M_{\rm A}
    + B[f_{\rm I}] M_{\rm A}^{3/2}
    = O(M_{\rm A}^{7/4}),
\end{equation}
where the functional $D$ is defined as
\begin{equation}
    D[f] = 1 + \dfrac{1}{2} \iint \dfrac{\partial_{p}f(q,p)}{p} {\rm d}q{\rm d}p,
\end{equation}
which coincides with the dispersion function with zero frequency
when $f$ does not depend on $q$.
For simplicity again, we assume that
the initial perturbation $g_{\rm I}= g(q,p,0)$ is even
with respect to both $q$ and $p$,
and can be expanded
into the Fourier series as
\begin{equation}
  \label{eq:gI}
  g_{\rm I}(q,p) = \sum_{n\geq 1} \tilde{g}_{\rm I}(n,p) \cos(nq).
\end{equation}
We note that, in this case, the contribution of the transient field 
$\mathcal{V}_{\rm T}$ to the asymptotic state is shown to vanish without approximation \eqref{eq:AVlasov}. 
See Appendix \ref{sec:vanish-T} for details.
Then, the functional $D[f_{\rm I}]$ is reduced to $D[f_{\rm S}]$,
and the functionals $A$ and $B$ are expressed by
\begin{equation}
  \label{eq:A}
    A[f_{\rm I}] = - \sum_{n\geq 1} \tilde{g}_{\rm I}(n,0) C_{n}
\end{equation}
and
\begin{equation}
  \label{eq:B}
    B[f_{\rm I}] = - f''_{\rm S}(0) C_{1} + O(\epsilon M_{\rm A}^{3/2}),
\end{equation}
where
\begin{equation}
    \label{eq:Cn}
    C_{n} = M_{\rm A}^{-1/2} \iint \cos(nq) \ave{\cos q}_{\mathcal{H}_{\rm A}} {\rm d}q{\rm d}p.
\end{equation}
We remark that $C_{n}$ does not depend on $M_{\rm A}$
due to the scaling of $\ave{\cos q}_{\mathcal{H}_{\rm A}}$,
and the constant values 
are numerically obtained as
\begin{equation}
    \label{eq:Cn-values}
    C_{0}=0, \quad
    C_{1}\simeq 5.169, \quad
    C_{2}\simeq 0.5360, \quad
    C_{3}\simeq -0.1043.
\end{equation}
Neglecting $O(M_{\rm A}^{7/4})$ terms in Eq. \eqref{eq:MA-expanded},
we have the solutions as
\begin{equation}
  \label{eq:approx-theory}
    \sqrt{M_{\rm A}} = 0,~ \dfrac{\sqrt{D^{2}-4AB}-D}{2B},
\end{equation}
where the second solution exists if and only if it is non-negative.

\section{Numerical tests of the rearrangement formula}
\label{sec:numerics}

\subsection{Two conditions to be tested}
\label{sec:two-conditions}

The rearrangement formula \eqref{eq:fA} predicts
the asymptotic value of order parameter, $M_{\rm A}$,
as the solutions to the self-consistent equation \eqref{eq:self-consistent},
or as approximation \eqref{eq:approx-theory}.
The zero solution corresponds to the strong Landau damping case,
and hence the non-zero solution, in which we are interested,
might be realized with the no Landau damping condition.
However, validity of this expectation is still not clear
since the rearrangement formula was successfully tested
for initial waterbag states satisfying the generalized virial condition
\cite{XL09,BMR11,ACRT14}.
Then, we will make competition between the no Landau damping condition
and a virial condition in Sec. \ref{sec:no-Landau-damping},
and will clarify that the former is more relevant in our setting.

The other condition comes from the Casimir invariants
of the Vlasov equation \eqref{eq:Vlasov},
where the invariants are
functionals of the form
\begin{equation}
  S[f] = \iint s(f) {\rm d}q{\rm d}p, \quad \text{$s$: $C^1$-function}.
\end{equation}
The rearrangement formula keeps all the Casimirs up to the linear order.
This fact is shown from the expansion
\begin{equation}
  \label{eq:linear-casimir-invariance}
  S[f_{\rm I}] - S[f_{\rm A}]
  = \iint s'(f_{\rm A}) \delta f {\rm d}\theta {\rm d}J
  + O((\delta f)^{2})
  = O((\delta f)^{2}),
\end{equation}
where $\delta f = f_{\rm I}-f_{\rm A}$.
The part of $s'(f_{\rm A})$ depends on $J$ only,
and $\ave{\delta f}_{\mathcal{H}_{\rm A}}=0$
from $\ave{f_{\rm I}}_{\mathcal{H}_{\rm A}}=f_{\rm A}$.
We again note that the angle-action variables $(\theta, J)$
associate with the asymptotic Hamiltonian $\mathcal{H}_{\rm A}$.
The invariance, however, does not hold in higher orders. Indeed, for
the Casimir
\begin{equation}
  \label{eq:S2}
  S_{2}[f] = \iint f^{2} {\rm d}q{\rm d}p=\iint f^{2} {\rm d}\theta {\rm d}J,
\end{equation}
we have the discrepancy as
\begin{equation}
  \label{eq:CasimirS2}
  \begin{split}
    S_{2}[f_{\rm I}] - S_{2}[f_{\rm A}]
    = \iint\ave{(\delta f)^{2}}_{\mathcal{H}_{\rm A}} {\rm d}q {\rm d}p,
  \end{split}
\end{equation}
which is not zero in general.
Therefore, the rearrangement formula may prefer initial states
with which the Casimir is not greatly modified.

The above two conditions are examined
by performing systematic numerical simulations
of the Vlasov equation \eqref{eq:Vlasov}.
We use the second-order semi-Lagrangian scheme \cite{PdB10}
with the cubic $B$ spline interpolations in each step.
Throughout this paper we use the truncated single-particle phase space
$(-\pi, \pi]\times [-4,4]$
and the time slice $\Delta t = 0.05$.
Asymptotic values are computed by taking
averages over the time interval $[500,1000]$
if no comment appears.
The phase space is divided into the grid of size $G\times G$,
which is called the grid size $G$.
In this section, we consider the HMF model without external field 
as in the previous section, 
and observe $M_{x}=\mathcal{M}_{x}[f]$.

\subsection{The no Landau damping condition}
\label{sec:no-Landau-damping}

The generalized virial condition represents quasistationarity
of a given waterbag initial state \cite{FPCB12},
and it is not straightforward to apply it for other initial states.
On the other hand, the proper virial condition is not useful
for spatially periodic systems.
Thus, for making competition with the no Landau damping condition,
we introduce another type of virial condition
with keeping the meaning of quasistationarity.

The proper virial condition is derived by differentiating
$P(t)=\sum_{j=1}^{N}p_{j}q_{j}/N$
and taking long-time average.
The periodic boundary condition of the HMF model
suggests to consider $Q(t)=\sum_{j=1}^{N}p_{j}\varphi(q_{j})/N$,
where $\varphi$ is an arbitrary smooth periodic function. 
Taking the limit $N\to\infty$, we replace the arithmetic mean
with the average over the distribution function $f$.
If $f_{\rm I}$ is stationary, we have the relation
\begin{equation}
    \ave{p^{2}\dfrac{{\rm d}\varphi}{{\rm d}q}(q)
      - \varphi(q) \dfrac{{\rm d}\mathcal{V}[f_{\rm I}]}{{\rm d}q}(q) }_{\rm I}=0,
\end{equation}
where $\ave{\bullet}_{\rm I} = \iint \bullet f_{\rm I} {\rm d}q {\rm d}p$.
Hereafter we put $\varphi(q)=\sin q$ which gives
\begin{equation}
    \label{eq:PVC}
    \ave{p^{2}\cos q
      - \sin q \dfrac{{\rm d}\mathcal{V}[f_{\rm I}]}{{\rm d}q}(q) }_{\rm I}=0.
\end{equation}
We refer to Eq. \eqref{eq:PVC} as the periodic virial condition.
We note that the above condition is equivalent with $\ddot{M}_{x}(0)=0$,
and $\dot{M}_{x}(0)=0$ is also satisfied for even $f_{\rm I}$
with respect to $p$. 
These vanishing derivatives imply that the periodic virial condition
represents quasistationarity in a short time interval.
The condition \eqref{eq:PVC} will be compared with
the no Landau damping condition,
which is explicitly written as
\begin{equation}
  \label{eq:NLD}
  D[f_{\rm S}] = 0
\end{equation}
since positive, negative and vanishing $D$ imply
that $f_{\rm S}$ is stable, unstable and marginal respectively.

For the competition, we prepare a family of initial states as
\begin{equation}
  \label{eq:MBTT}
  f_{\rm I}(q,p;T_{0},T_{1}) = f_{\rm MB}(p;T_{0}) + \epsilon f_{\rm MB}(p;T_{1}) \cos q,
\end{equation}
where $f_{\rm MB}$ denotes the Maxwell-Boltzmann distribution
\begin{equation}
    f_{\rm MB} (p;T) =\dfrac{1}{2\pi \sqrt{2\pi T}} e^{-p^2/2T}.
\end{equation}
The unperturbed part gives
\begin{equation}
    D[f_{\rm MB}] = 1 - \dfrac{1}{2T_{0}},
\end{equation}
and the no Landau damping condition $D=0$
is realized at
the critical temperature $T_{0}=T_{\rm c}=1/2$
of the second order phase transition \cite{IK93,MA95}. 
The Maxwell-Boltzmann is stable for $T>T_{\rm c}$.
On the other hand, the periodic virial condition \eqref{eq:PVC}
is realized at $T_{1}=T_{\rm c}=1/2$.
The family \eqref{eq:MBTT},
therefore, can exclusively satisfy one of the two conditions as follows:
\begin{description}
    \vspace{-1mm}
      \item[Case 1] $f_{\rm I}(q,p; T_{\rm c},T)$ with $T > T_{\rm c}$ :
    The unperturbed term $f_{\rm MB}(p;T_{\rm c})$ satisfies
    the no Landau damping condition \eqref{eq:NLD},
    but $f_{\rm I}$ breaks the periodic virial condition
    \eqref{eq:PVC}.
    \vspace{-2mm}
      \item[Case 2] $f_{\rm I}(q,p; T, T_{\rm c})$ with $T> T_{\rm c}$ :
    $f_{\rm I}$ satisfies the periodic virial condition \eqref{eq:PVC},
    but the unperturbed term $f_{\rm MB}(p;T)$ breaks
    the no Landau damping condition \eqref{eq:NLD}.
\end{description}

\begin{figure}[tb]
    \begin{center}
	\includegraphics[width=8.5cm]{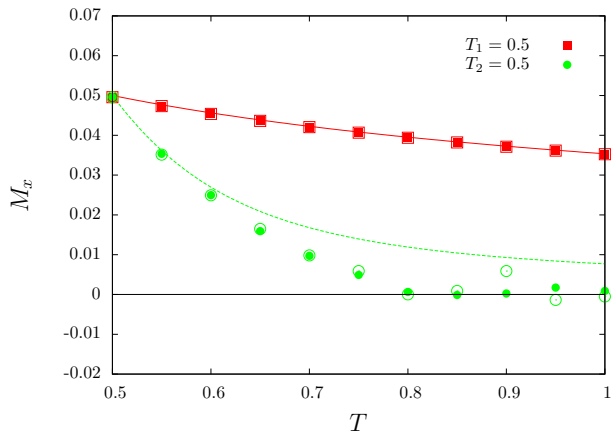}
	\caption{
          (color online)
          Asymptotic values of $M_{x}$ for the initial distributions
          \eqref{eq:MBTT}.
          The red squares are for {\bf Case 1},
          and the green circles are for {\bf Case 2}.
          The open and filled symbols are computed
          with the grid sizes 
          $G=256$ and $512$ respectively.
          $\epsilon=0.1$.
          The red solid and the green dashed lines
          are from the approximated theory \eqref{eq:approx-theory}
          for {\bf Case 1} and {\bf Case 2} respectively.
	}
	\label{fig:DoubleGaussian}
    \end{center}
\end{figure}

As shown in Fig. \ref{fig:DoubleGaussian},
the rearrangement formula gives precise prediction
in {\bf Case 1} for all $T>T_{\rm c}$ and 
in {\bf Case 2} for $T$ close to $T_{\rm c}$. 
In {\bf Case 2}, the agreement between the rearrangement formula
and numerics becomes worse as $T$ increases, that is, 
the Landau damping rate gets larger,
though the periodic virial condition holds.
From the numerical observation,
we conclude that the no Landau damping condition is more relevant
than the periodic virial condition for the perturbed
Maxwell-Boltzmann states \eqref{eq:MBTT}.

\subsection{Casimir invariance}
\label{sec:Casimir}

The previous work \cite{SO14} uses the initial perturbation $g_{\rm I}$
having only the first Fourier mode
with respect to the position $q$ (see Eq. \eqref{eq:gI}),
and shows that the rearrangement formula gives precise predictions
at the critical point even for rather large perturbation.
However, any Fourier modes can contribute to the asymptotic value
of order parameter $M_{\rm A}$ through mode couplings.
We will reveal that initial perturbation is also restrictive
by adding the second Fourier mode to it,
and will qualitatively explain discrepancy between the theory 
and numerics from the view point of the Casimir invariance.

We prepare the initial state as
\begin{equation}
  \label{eq:test-IC}
    f_{\rm I}(q,p)
    = f_{\rm MB}(p;T_{\rm c})
      \sum_{k=0}^{2} \epsilon_{k}\cos kq,
\end{equation}
where $\epsilon_{0}=1$.
We call the term $\epsilon_{k}\cos kq$ the $k$-th mode.
We set the stationary state as $f_{\rm S}(p)=f_{\rm MB}(p;T_{\rm c})$
for an independent test of the no Landau damping condition.
The first mode is included to escape from $M_{x}=0$.

\begin{figure}[tb]
    \begin{center}
	\includegraphics[width=8.5cm]{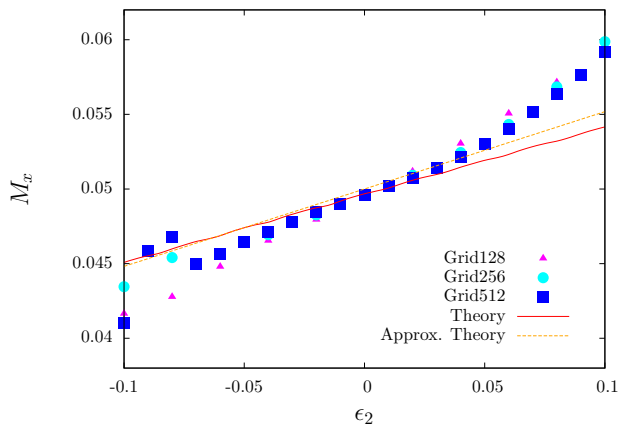}
	\caption{
          (colour online) $\epsilon_{2}$ dependence of $M_{x}$
          with the fixed $\epsilon_{1}=0.1$
          for the initial state \eqref{eq:test-IC}.
          The red solid line represents the full theory \eqref{eq:self-consistent},
          the orange dashed line approximated theory \eqref{eq:approx-theory},
          and points are from numerics.
          The sizes of grid are $G=128$ (purple triangles),
          $256$ (light blue circles) and $512$ (blue squares).
	}
	\label{fig:01epsilon1}
    \end{center}
\end{figure}

For a fixed value of $\epsilon_{1}=0.1$,
we show $\epsilon_{2}$ dependence of $M_{x}$ in Fig. \ref{fig:01epsilon1}.
The theoretical prediction is in good agreement with numerics
for small $|\epsilon_{2}|$, but discrepancy tends to grow
for large $|\epsilon_{2}|$.
Moreover, in large $\epsilon_{2}$ region,
the theoretical prediction is smaller than numerics,
while the Landau damping mechanism provides inverse result.
Existence of the second mode is, therefore, an independent mechanism
to yield discrepancy between the theory and numerics.
We remark that there is a non-regular dependence on $\epsilon_{2}$
around $\epsilon_{2}\simeq -0.08$,
but mechanism of this dependence is not clear yet.

\begin{figure}[tb]
  \begin{center}
    \includegraphics[width=8.5cm]{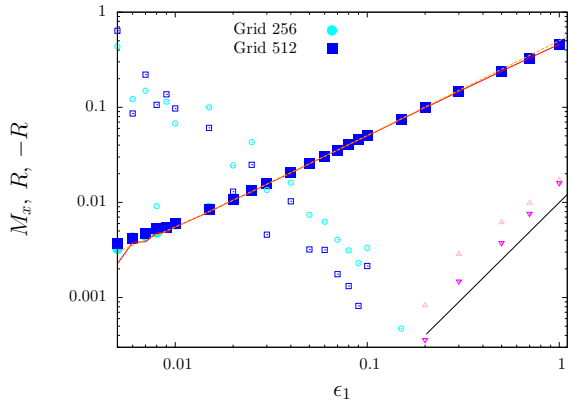}
    \caption{
      (color online)
      $\epsilon_{1}$ dependence of $M_{x}$ with the fixed $\epsilon_{2}=0.01$
      for the initial state \eqref{eq:test-IC}.
      The solid red line represents the full theory \eqref{eq:self-consistent},
      the orange dashed line the approximated theory \eqref{eq:approx-theory},
      and filled points are from numerics.
      The grid sizes are $G=256$ (light blue circles)
      and $512$ (blue squares).
      Open symbols represent relative errors with the full theory:
      $R$ for $G=256$ (light blue circles) and $512$ (blue squares),
      and $-R$ for $G=256$ (pink triangles) and $512$ (purple inverse triangles).
      The black solid line is guide for eyes,
      and has slope $2$.
    }
	\label{fig:001epsilon2}
    \end{center}
\end{figure}

Looking at Fig. \ref{fig:01epsilon1},
we expect that $|\epsilon_{2}|$ must be much smaller than $|\epsilon_{1}|$.
The above expectation is confirmed by varying $\epsilon_{1}$
for a fixed value of $\epsilon_{2}=0.01$.
Values of the asymptotic magnetization $M_{x}$
are reported in Fig. \ref{fig:001epsilon2}
with the relative error defined by
\begin{equation}
  R = \dfrac{M_{\rm numerics}-M_{\rm theory}}{M_{\rm theory}}
\end{equation}
where $M_{\rm theory}$ and $M_{\rm numerics}$ 
are respectively obtained theoretically \eqref{eq:self-consistent}
and numerically.
In the large $\epsilon_{1}$ region,
the minus of relative error grows as $\epsilon_{1}$ gets large.
This growth of the relative error of $O(\epsilon_{1}^{2})$
might be rather natural since we omitted
$O(\epsilon^{2})$ terms in Eq. \eqref{eq:TLVlasov}.
Interesting observations are that the relative error
changes the sign around the minimum point,
and grows even $\epsilon_{1}$ decreases.

Let us discuss mechanism of the $\epsilon_{1}$ dependence of
the relative error
from the view point of invariance of the Casimir $S_{2}$ \eqref{eq:S2}.
The initial value of $S_{2}$ is computed as
\begin{equation}
    \label{eq:S2fI}
    S_{2}[f_{\rm I}] = \iint f_{\rm I}(q,p)^{2}{\rm d}q{\rm d}p
    = \dfrac{1}{4\pi\sqrt{\pi T_{\rm c}}}
    \left( 1 + \dfrac{\epsilon_{1}^{2}}{2} + \dfrac{\epsilon_{2}^{2}}{2} \right).
\end{equation}
On the other hand, under some phenomenological assumptions,
we approximate $S_{2}[f_{\rm A}]$ as
\begin{equation}
  \label{eq:S2fA}
  S_{2}[f_{\rm A}]  
  \simeq \dfrac{1}{4\pi\sqrt{\pi T_{\rm c}}} \left(
    1 + \dfrac{\epsilon_{1}^{2}}{2} + 2c_{2} \epsilon_{1}\epsilon_{2}
    + c_{2}^{2}\epsilon_{2}^{2}
  \right),
\end{equation}
where $c_{2}=C_{2}/C_{1}\simeq 0.1$
from the values of $C_{1}$ and $C_{2}$, \eqref{eq:Cn-values}.
See the Appendix \ref{sec:casimir-computation}
for deriving the approximation \eqref{eq:S2fA}.
Comparing the asymptotic value \eqref{eq:S2fA} with the initial value
\eqref{eq:S2fI}, we find that invariance of the Casimir $S_{2}$
is realized for $\epsilon_{2}$ satisfying
\begin{equation}
  \epsilon_{2}^{2} = 4c_{2}\epsilon_{1}\epsilon_{2}
  + 2c_{2}^{2}\epsilon_{2}^{2},
\end{equation}
which is approximately solved by
\begin{equation}
  \label{eq:relation-epsilon12}
  \epsilon_{2} \simeq 4c_{2} \epsilon_{1}
  \simeq 0.4 \epsilon_{1}.
\end{equation}
The relation \eqref{eq:relation-epsilon12} qualitatively
explains the minimum point of the relative error in Fig.\ref{fig:001epsilon2}.

The estimation of $S_{2}[f_{\rm A}]$, \eqref{eq:S2fA},
also qualitatively explains underestimation by the theory
for large $\epsilon_{2}/\epsilon_{1}$.
Suppose $\epsilon_{2}\simeq\epsilon_{1}$.
In this case $S_{2}[f_{\rm I}]$ is larger than $S_{2}[f_{\rm A}]$
due to the factor $c_{2}=C_{2}/C_{1}\simeq 0.1$.
In the full Vlasov system, $S_{2}$ is conserved
and hence the lost part $S_{2}[f_{\rm I}]-S_{2}[f_{\rm A}]$
must be covered by, for instance,
increasing amplitude of the first Fourier mode
relating to $M_{\rm A}$.
Then, underestimation by the theory possibly occurs.

\section{Response theory to the external field}
\label{eq:response-theory}
We have dealt with the HMF model without external field  
in Secs. \ref{sec:rearrangement-formula} and \ref{sec:numerics}. 
In this section, we consider response to the non-zero external field
and the critical exponents $\gamma_{\pm}$ and $\delta$
defined as
\begin{equation}
  M_{\rm A}-M_{\rm I} \propto |T-T_{\rm c}|^{-\gamma_{\pm}} h,
  \quad
  M_{\rm A} \propto h^{1/\delta}
\end{equation}
in the limit of $h\to 0$, where $h$ is strength
of the external field, $M_{\rm I}$ the initial order parameter,
and $\gamma_{+}$ and $\gamma_{-}$ are defined
in high- and low-temperature sides respectively.
We note that the considering family of states may be
thermal equilibrium states or QSSs,
and a certain parameter plays the role of temperature in the latter QSS case.
The rearrangement formula gives
the non-classical critical exponents of $\gamma_{+}=1$,
$\gamma_{-}=1/4$ \cite{OPY14} and $\delta=3/2$ \cite{SO14}
in a wide class of 1D Vlasov dynamics with the periodic boundary condition
\cite{OY15},
while statistical mechanics gives $\gamma_{+}=\gamma_{-}=1$
and $\delta=3$.

We numerically found that the rearrangement formula
requires two conditions to be satisfied:
One is the no Landau damping condition and
the other comes from the Casimir invariance.
These conditions restrict applicable initial states for
the rearrangement formula.
However, the conditions reinforce validity
of use of the rearrangement formula
as the response theory to the external field,
which turns on at the initial time and goes to be constant asymptotically,
since this setting satisfies both the two conditions
as discussed in the following.

Performing the Laplace transform of the external field
we get a pole at the origin of the Laplace space
(the complex frequency plane),
and the pole provides the asymptotically surviving response \cite{AP12,OY12}.
We hence conjecture that this pole effectively restores
the no Landau damping condition even if the unperturbed stationary state
breaks the condition.
Moreover, for the critical exponents,
the interesting reference states are close to the critical state
which satisfies the no Landau damping condition.

It is not hard to see that the second condition,
the Casimir invariance, is satisfied
for small $h$.
As shown in Eq. \eqref{eq:linear-casimir-invariance},
invariance of all the Casimirs is satisfied
by the rearrangement theory up to the linear order.
The critical exponents are defined in the limit of small external field,
and hence the Casimir invariance is not an obstacle for computing them.

\section{Summary and Discussions}
\label{sec:summary}
We discussed conditions to use the rearrangement formula
around spatially homogeneous stable stationary states
in the HMF model, and numerically derived two conditions:
One is for the stable stationary reference state,
and the other is for the whole initial state. 
The former is the no Landau damping condition,
which was previously suggested \cite{SO14}.
We compared this condition with a virial condition,
which we called the periodic virial condition,
and numerically clarified that the no Landau damping condition
is more crucial than the periodic virial condition.
The latter comes from the Casimir invariance:
The theory prefers initial perturbed states
which keep the Casimirs well.
Breaking the former and the latter,
theoretical predictions tend to overestimate and underestimate
respectively, and hence we may conclude
that the two conditions are independent.

Due to the conditions, 
the rearrangement formula is restrictive
for using as a nonequilibrium statistical mechanics.
Nevertheless, the theory is useful as a response theory
to the external field saturating to a small constant asymptotically,
since the conditions are satisfied in such a situation.
In particular, the conditions validate to compute the critical exponents
in use of the rearrangement formula.

Another important benefit of the present work is
that the conditions suggest a direction for improving the rearrangement formula:
The theory could be improved by inputting the Landau damping
and the Casimir invariants.
For instance, we expect that nonlinear trapping plays an important role
to form a magnetized asymptotic state,
and an improved theory may be derived by considering the competition
between the linear Landau damping and the nonlinear trapping
as discussed for forming traveling small clusters \cite{JB09}.
Such an improvement is interesting and worthwhile
for constructing a nonequilibrium statistical mechanics,
but remains as a future work.

  A similar formula with the rearrangement formula
  has been also derived by de Buyl {\it et al.} \cite{PdB13}
  for a small system of $O(\epsilon)$
  contacting with a huge bath of $O(1)$
  through long-range interactions.
  In this setting, magnetization in the huge bath plays the role of
  external field for the system,
  and the rearrangement formula possibly provides good predictions
  as discussed in the present article,
  if the bath is huge enough and static accordingly.
  We remark that the system is driven by the bath magnetization only,
  and no self-consistent condition is needed for the system magnetization,
  since the latter is small enough and can be omitted.

This article dealt with the HMF model only, but,
from physical mechanism leading the two conditions,
one may expect that generic systems having long-range interactions
share the two conditions.
Examinations for several systems remain to be done.
We discussed initial states around stable stationary states,
but the rearrangement formula has been successfully used
in 3D self gravitating systems with watarbag initial conditions,
which satisfy the virial condition \cite{BRPL14}.
It also remains to reveal a relation
between the two types of initial states,
which are perturbed stable stationary states and waterbag states.

We end this article by mentioning the discussion on
parametric resonance for initial states which are
neither perturbed stable stationary states nor waterbag states
satisfying the virial condition \cite{BRPL14}.
We have discussed on the discrepancy induced with the
higher Fourier modes based on the Casimir invariants.
On the other hand, there is another explanation based on the
parametric resonance
induced by the higher moments \cite{BRPL14}.
Clarifying the relation between the two explanations remains
as another future work.

\acknowledgements
Y.Y.Y. acknowledges the support of JSPS KAKENHI Grant Number 23560069.
S.O. acknowledges the support of Grant-in-Aid for JSPS Fellows Grant Number 254728. 

\appendix

\section{Explicit form of $\ave{\bullet}_{\mathcal{H}_{\rm A}}$}
\label{sec:explicit-average}

The angle-action variables $(\theta,J)$ are obtained by
\begin{equation}
  \theta = \dfracp{W}{J}(q,J),
  \quad
  J = \dfrac{1}{2\pi} \oint p {\rm d}q, 
\end{equation}
where the integral is performed along a periodic orbit,
and the generating function $W$ is
\begin{equation}
  W(q,J) = \int_{0}^{q} p(q',J) {\rm d}q'.
\end{equation}
To express these variables for the asymptotic effective Hamiltonian
system 
\begin{equation}
  \mathcal{H}_{\rm A} = \dfrac{p^{2}}{2} - M_{\rm A}\cos q,
\end{equation}
we use the variable $k$ defined as
\begin{equation}
	k \equiv \sqrt{\frac{ \mathcal{H}_{\rm A} + M_{\rm A}}{2 M_{\rm A}}},
\end{equation}
and the Legendre elliptic integrals of the first and the second kinds
respectively defined by
\begin{equation}
	\begin{split}
	  F(\phi,k) &= \int_{0}^{\phi} \dfrac{{\rm d}x}{\sqrt{1-k^{2}\sin^{2}x}}, \\
 	  E(\phi,k) &= \int_{0}^{\phi} \sqrt{1-k^{2}\sin^{2}x} {\rm d}x.
	\end{split}
\end{equation}
These integrals induce the complete elliptic integrals
of the first and the second kinds respectively defined as
\begin{equation}
  K(k) = F(\pi/2,k), \quad E(k) = E(\pi/2,k).
\end{equation}
The action and the angle variables are then expressed in the forms
\begin{equation}
  J = 
  \begin{cases}
    \dfrac{8\sqrt{M_{\rm A}}}{\pi} [ E(k)-(1-k^{2})K(k)] , & \quad k<1 \\ 
    \dfrac{4\sqrt{M_{\rm A}}k}{\pi} E(1/k), & \quad k>1 \\ 
  \end{cases}
\end{equation}
and
\begin{equation}
  \theta = 
  \begin{cases}
    \frac{\pi}{2} F(Q,k)/K(k), &\quad p \geq 0,~ k<1 \\ 
    \frac{\pi}{2} \left(2- F(Q,k)/K(k)\right), &\quad p<0, ~ k<1\\
    \pi {\rm sgn}(p) F(Q,1/k)/K(1/k),  & \quad k>1, \\
  \end{cases} 
\end{equation}
where $Q$ is defined as $k\sin Q=\sin(q/2)$ for $k<1$
and as $Q=q/2$ for $k>1$. See Ref. \cite{JB10} for details.

Using the variable transforms from $\theta$ to $Q$,
for an observable $B(q,p)$ even with respect to $p$,
we can write the average of $B(q,p)$ over an iso-$\mathcal{H}_{\rm A}$ curve as
\begin{equation}
  \ave{B}_{\mathcal{H}_{\rm A}} = 
  \begin{cases}
    \dfrac{1}{2K(k)} \displaystyle{\int_{-\pi/2}^{\pi/2} \dfrac{B(q,p)}{\sqrt{1-k^{2}\sin^{2}Q}}} {\rm d}Q, & \quad k<1 \\
    \dfrac{1}{2K(1/k)} \displaystyle{\int_{-\pi/2}^{\pi/2} \dfrac{B(q,p)}{\sqrt{1-k^{-2}\sin^{2}Q}}} {\rm d}Q, & \quad k>1 \\
  \end{cases}
\end{equation}
where $B(q,p)$ must be transformed to a function of $(Q,k)$.
The average $\ave{B}_{\mathcal{H}_{\rm A}}$ is obtained as a function of $k$.

For a Hamiltonian system $\mathcal{H}(q,p)$,
there is another expression of the average over iso-energy curves as
\begin{equation}
  \ave{B}_{\delta}
  = \dfrac{\displaystyle{\iint \delta(\mathcal{H}(q,p)-E) B(q,p) {\rm d}q{\rm d}p}}
  {\displaystyle{\iint \delta(\mathcal{H}(q,p)-E) {\rm d}q{\rm d}p}}.
\end{equation}
This expression has been applied to the HMF model \cite{BMR11, ACRT14}
and to the 3D self-gravitating system \cite{BRPL14}.
We can show the equality $\ave{B}_{H}=\ave{B}_{\delta}$ for the 1D case
in each region of phase space where we can construct the inverse
function of the Hamiltonian $H(J)$.
Using the relation ${\rm d}q{\rm d}p={\rm d}\theta {\rm d}J$,
and the variable change $x=H(J)$, we can modify $\ave{B}_{\delta}$ as
\begin{equation}
  \begin{split}
    \ave{B}_{\delta}
    & = \dfrac{\displaystyle{\iint \delta(x-E) B(\theta,H^{-1}(x)) {\rm d}\theta \dfrac{{\rm d}x}{\Omega(\mathcal{H}^{-1}(x))}}}
    {\displaystyle{\iint \delta(x-E) {\rm d}\theta \dfrac{{\rm d}x}{\Omega(\mathcal{H}^{-1}(x))}}} \\
    & = \dfrac{1}{2\pi} \int B(\theta,\mathcal{H}^{-1}(E)) {\rm d}\theta
    = \ave{B}_{\mathcal{H}},
  \end{split}
\end{equation}
where $\Omega(J)=({\rm d}\mathcal{H}/{\rm d}J)$, we used the fact that 
$\Omega(J) > 0$ except for $J$ corresponding to the separatrix,
and we denoted the observable $B$ as $B(\theta,J)$
even in the angle-action coordinate for simplicity of notation. 
The expression $\ave{B}_{\delta}$ might be useful
when deriving the angle-action variables is hard.

\section{Derivation of the rearrangement formula without omitting 
the term including T-field $\mathcal{V}_{\rm T}$}
\label{sec:vanish-T}

Precisely, the asymptotic part $f_{\rm A}$ is constructed by the two terms 
of so-called 
the O'Neil term $f_{\rm O}$ and the Landau term $f_{\rm L}$ 
defined by 
\begin{equation}
	\begin{split}
		f_{\rm O} &= e^{-t\{\mathcal{H}_{\rm A}, \bullet\}} f_{\rm I},  \\
		f_{\rm L} &= -\int_0^t  e^{-(t-s)\{\mathcal{H}_{\rm A}, \bullet\}} \{\mathcal{V}_{\rm T}(s), f_{\rm I} \} {\rm d} s,
	\end{split}
\end{equation}
respectively \cite{CL03,CL09}. 
By use of them, 
the solution to the T-linearized equation \eqref{eq:TLVlasov} indeed written as 
$f_{\rm TL}(q,p,t) = f_{\rm O}(q,p,t) + \epsilon f_{\rm L}(q,p,t)$. 
The O'Neil term gives the expression
$\ave{f_{\rm I}}_{\mathcal{H}_{\rm A}}$ in the limit $t \to \infty$. 
The Landau term comes from the neglected 
term of $\epsilon\{\mathcal{V}_{\rm T}, f_{\rm I}\}$
(see Eq.\eqref{eq:TLVlasov}),
and is neglected since it has no contribution
to the asymptotic Hamiltonian $\mathcal{H}_{\rm A}$ \cite{SO14}. 
Meanwhile, 
it has not been shown that contribution of the Landau term to the asymptotic 
distribution vanishes or not. 
We show that the Landau term completely vanish in the limit of $t \to \infty$, when the initial state $f_{\rm I}(q,p)$ 
is even with respect to both $q$ and $p$, that is, 
$f_{\rm I}(q,p) = f_{\rm I}(-q,p) =f_{\rm I}(-q,-p) = f_{\rm I}(q,-p)$.
The initial conditions dealt in this paper have this symmetry.

Let us show $f(q,p,t) = f(-q,-p,t)$ for $t \geq 0$ if $f_{\rm I}(q,p)= f_{\rm I}(-q,-p)$ at initial. 
Changing variables by $(q,p) \mapsto (-q,-p)$, 
it is easy to show that $f(-q,-p,t)$ is also a solution to the Vlasov equation with the initial condition 
$f_{\rm I}(q,p)$. 
It is, then, shown that $f(q,p,t) = f(-q,-p,t)$, 
due to the existence and uniqueness of solution to the Vasov equation \cite{Spohn}. 
The fact $\mathcal{M}_y[f](t) = 0$ is immediately shown, 
and it is reasonable to consider that the asymptotic Hamiltonian 
can be given in the form
\begin{equation}
  \mathcal{H}_{\rm A}(q,p) = p^2/2 - M_{\rm A} \cos q ,
\end{equation}
which says $\mathcal{M}_{y}[f_{\rm A}]=0$.
Thus, the definition of transient part $g_{\rm T}$ [Eq. \eqref{eq:ATP}],
induces that $\mathcal{M}_{y}[g_{\rm T}](t)=0$.
 
The asymptotic form of the Landau term is written in the form
\begin{equation}
  \label{eq:Landauterm}
  \begin{split}
    \lim_{t \to \infty} f_{\rm L}
    & =
    \left\langle \sin q \frac{\partial f_{\rm I}}{\partial p} \right\rangle_{\mathcal{H}_{\rm A}}
    \int_{0}^{\infty} \mathcal{M}_{x}[g_{\rm T}](t) {\rm d}t\\
    & - \left\langle \cos q \frac{\partial f_{\rm I}}{\partial p}\right\rangle_{\mathcal{H}_{\rm A}}
  \int_{0}^{\infty} \mathcal{M}_{y}[g_{\rm T}](t) {\rm d}t.
  \end{split}
\end{equation}
The symmetry for $q\to -q$ of $f_{\rm I}$ and of $\mathcal{H}_{\rm A}$
vanishes the first term of the right-hand-side,
and the fact $\mathcal{M}_{y}[g_{\rm T}](t)=0$ eliminates the second.
We, therefore, conclude $\lim_{t\to\infty} f_{\rm L}=0$.
This procedure can be applied to more general systems
that we have dealt in Ref. \cite{OY15}.

\section{Asymptotic value of the Casimir $S_{2}$}
\label{sec:casimir-computation}

The Jeans theorem \cite{JHJ15} states
that a state is stationary if and only if it depends on $(q,p)$
solely through the first integrals,
in our case, the effective Hamiltonian.
The asymptotic stationary state is, therefore, expressed as
\begin{equation}
    f_{\rm A}(q,p) = F_{\rm A}(p^{2}/2-M_{\rm A}\cos q).
\end{equation}
Now $M_{\rm A}$ is assumed to be small and we
further assume that $F_{\rm A}$ accepts the Taylor expansion
\begin{equation}
    \label{eq:FA}
    \begin{split}
        F_{\rm A}(p^{2}/2-M_{\rm A}\cos q)
        & = F_{\rm A}(p^{2}/2) - M_{\rm A}\cos q F'_{\rm A}(p^{2}/2) \\
        & + \dfrac{M_{\rm A}^{2}}{2}\cos^{2}q F''_{\rm A}(p^{2}/2) + \cdots.
    \end{split}
\end{equation}
Validity of this assumption
is not obvious due to the averaging procedure $f_{\rm A} = \langle f_{\rm I} \rangle_{\mathcal{H}_{\rm A}}$
which makes a cusp for the function
$\langle \cos q \rangle_{\mathcal{H}_{\rm A}}$
at the separatrix energy \cite{SO14} for instance.
However, it
helps us to discuss on the Casimir $S_{2}$ qualitatively.

We assume that $F_{\rm A}(p^{2}/2)$ is close to the unperturbed
Maxwell-Boltzmann distribution $f_{\rm MB}(p;T_{\rm c})=F_{\rm MB}(p^{2}/2)$,
where $F_{\rm MB}(E)\propto e^{-E/T_{\rm c}}$, and write it as
\begin{equation}
  F_{\rm A}(p^{2}/2) = F_{\rm MB}(p^{2}/2) + \epsilon G(p^{2}/2)
\end{equation}
with a small parameter $\epsilon$ of $O(\epsilon_{1})$ or $O(\epsilon_{2})$.
Remembering that $M_{\rm A}$ is of $O(\epsilon)$,
omitting $O(\epsilon^{3})$ and using $F_{\rm MB}'=-F_{\rm MB}/T_{\rm c}$, we have
\begin{equation}
  \begin{split}
    f_{\rm A}
    & \simeq \left( 1 + \dfrac{M_{\rm A}^{2}}{4T_{\rm c}^{2}} \right)
    F_{\rm MB} + \epsilon G \\
    & + \left( \dfrac{M_{\rm A}}{T_{\rm c}} F_{\rm MB} - \epsilon M_{\rm A} G' \right)
    \cos q
    + \dfrac{M_{\rm A}^{2}}{4T_{\rm c}^{2}} F_{\rm MB} \cos 2q.
  \end{split}
\end{equation}
Straightforward computations give
\begin{equation}
  S_{2}[f_{\rm A}]
  \simeq \dfrac{1}{4\pi\sqrt{\pi T_{\rm c}}}
  \left( 1 + \dfrac{M_{\rm A}^{2}}{T_{\rm c}^{2}}
    + 2 \epsilon \dfrac{\int F_{\rm MB}G dp}{\int F_{\rm MB}^{2} dp}
    + \epsilon^{2} \dfrac{\int G^{2}dp}{\int F_{\rm MB}^{2} dp}
  \right)
\end{equation}
by omitting $O(\epsilon^{3})$.
Introducing $c_{2}=C_{2}/C_{1}$ and remembering that
\begin{equation}
  M_{\rm A} = - \dfrac{A}{B}
  = T_{\rm c}\left( \epsilon_{1} + c_{2}\epsilon_{2} + \cdots \right)
\end{equation}
at the critical point $T_{\rm c}$
from Eqs. \eqref{eq:A}, \eqref{eq:B} and \eqref{eq:approx-theory},
the asymptotic value is rewritten as
\begin{equation}
  \label{eq:S2fA-approximate}
  \begin{split}
    & S_{2}[f_{\rm A}]
    \simeq \dfrac{1}{4\pi\sqrt{\pi T_{\rm c}}}
    \left( 1 + \dfrac{(\epsilon_{1}+c_{2}\epsilon_{2})^{2}}{2} \right ) \\
    & + \dfrac{1}{4\pi\sqrt{\pi T_{\rm c}}}
    \left( \dfrac{(\epsilon_{1}+c_{2}\epsilon_{2})^{2}}{2}
    + 2 \epsilon \dfrac{\int F_{\rm MB}G dp}{\int F_{\rm MB}^{2} dp}
    + \epsilon^{2} \dfrac{\int G^{2}dp}{\int F_{\rm MB}^{2} dp}
  \right).
  \end{split}
\end{equation}

It has been numerically reported that
the rearrangement formula gives precise predictions
for $\epsilon_{2}=0$ \cite{SO14}.
Thus, comparing \eqref{eq:S2fA-approximate} with \eqref{eq:S2fI},
we assume that the equality
\begin{equation}
  \label{eq:smallness-assumption}
  \epsilon_{1}^{2} + 4\epsilon\dfrac{\int F_{\rm MB}G dp}{\int F_{\rm MB}^{2} dp}
  + 2\epsilon^{2} \dfrac{\int G^{2}dp}{\int F_{\rm MB}^{2} dp}
  = O(\epsilon^{3})
\end{equation}
holds for $\epsilon_{2}=0$ and for small $\epsilon_{2}$.
This assumption induces the asymptotic value of Eq. \eqref{eq:S2fA}.

\end{document}